\documentclass[aps,prl,twocolumn,showpacs,byrevtex]{revtex4}

\usepackage{times,mathptm}% for making `pdf' file
\usepackage[dvips]{graphicx,color}% for inclusion of graphics file(s)
\usepackage{titlesec}

\begin{document}

\title{Dirac Fermions in Antiferromagnetic FeSn Kagome Lattices with Combined Space Inversion and Time Reversal Symmetry}
\author{Zhiyong Lin,$^{1,\dagger}$ Chongze Wang,$^{2,\dagger}$ Pengdong Wang,$^{3,\dagger}$ Seho Yi,$^{2,\dagger}$ Lin Li,$^{1,*}$ Qiang Zhang,$^{1}$ Yifan Wang,$^{1}$ Zhongyi Wang,$^{1}$ Hao Huang,$^{1}$ Yan Sun,$^{4}$ Yaobo Huang,$^{5}$ Dawei Shen,$^{6}$ Donglai Feng,$^{1,7}$ Zhe Sun,$^{3,*}$ Jun-Hyung Cho,$^{2,*}$ Changgan Zeng,$^{1,*}$ and Zhenyu Zhang$^1$}
\affiliation{$^1$International Center for Quantum Design of Functional Materials, Hefei National Laboratory for Physical Sciences at the Microscale, CAS Key Laboratory of Strongly Coupled Quantum Matter Physics, Department of Physics, and Synergetic Innovation Center of Quantum Information $\&$ Quantum Physics, University of Science and Technology of China, Hefei, Anhui 230026, China \\
$^2$Department of Physics, Research Institute for Natural Science, and HYU-HPSTAR-CIS High Pressure Research Center, Hanyang University, 222 Wangsimni-ro, Seongdong-Ku, Seoul 04763, Republic of Korea \\
$^3$National Synchrotron Radiation Laboratory, University of Science and Technology of China, Hefei, Anhui 230029, China \\
$^4$Max Planck Institute for Chemical Physics of Solid, Dresden, Germany \\
$^5$Shanghai Synchrotron Radiation Facility, Shanghai Institute of Applied Physics, Chinese Academy of Sciences, Shanghai 201204, China \\
$^6$State Key Laboratory of Functional Materials for Informatics and Center for Excellence in Superconducting Electronics, Shanghai Institute of Microsystem and Information Technology, Chinese Academy of Sciences, Shanghai 200050, China \\
$^7$State Key Laboratory of Surface Physics, Department of Physics, and Advanced Materials Laboratory, Fudan University, Shanghai 200438, China
}

\date{\today}

\begin{abstract}
Symmetry principles play a critical role in formulating the fundamental laws of nature, with a large number of symmetry-protected topological states identified in recent studies of quantum materials. As compelling examples, massless Dirac fermions are jointly protected by the space inversion symmetry \emph{P} and time reversal symmetry \emph{T} supplemented by additional crystalline symmetry, while evolving into Weyl fermions when either \emph{P} or \emph{T} is broken. Here, based on first-principles calculations, we reveal that massless Dirac fermions are present in a layered FeSn crystal containing antiferromagnetically coupled ferromagnetic Fe kagome layers, where each of the \emph{P} and \emph{T} symmetries is individually broken but the combined \emph{PT} symmetry is preserved. These stable Dirac fermions protected by the combined \emph{PT} symmetry with additional non-symmorphic $S_{\rm{2z}}$ symmetry can be transformed to either massless/massive Weyl or massive Dirac fermions by breaking the \emph{PT} or $S_{\rm{2z}}$ symmetry. Our angle-resolved photoemission spectroscopy experiments indeed observed the Dirac states in the bulk and two-dimensional Weyl-like states at the surface. The present study substantially enriches our fundamental understanding of the intricate connections between symmetries and topologies of matter, especially with the spin degree of freedom playing a vital role.
\end{abstract}

\maketitle

Topological materials have been extended to include gapless systems that are characterized by the nontrivial topology of bulk bands and its associated robust surface states. Such topological semimetals host electronic structures with linear band-contact nodes or lines \cite{Armitage18,Wang13,Wang12,Liu14,Liu14_2,Wan11,Wu16}. Existing studies of three-dimensional (3D) Dirac semimetals have been primarily restricted to nonmagnetic materials \cite{Armitage18,Wang13,Wang12,Liu14,Liu14_2,Neupane14,Borisenko14} with both \emph{T} and \emph{P} symmetries, where the electronic bands are doubly degenerate at each momentum and two of such doubly degenerate bands accidentally cross to form a Dirac point. This four-fold degenerate Dirac point can be stable against spin-orbit coupling (SOC) under additional crystalline symmetry such as glide mirror symmetry or screw rotation symmetry \cite{Armitage18,Yang14,Yang17}. Recently, Shou-Cheng Zhang and his colleagues presented the conceptually intriguing proposal that Dirac fermions can also be hosted even in antiferromagnetic (AFM) materials, where both \emph{T} and \emph{P} are broken but their combination \emph{PT} is preserved \cite{Tang16}. However, experimental realization of the Dirac fermions in realistic AFM materials remains to be accomplished. Contrasting with such 3D Dirac fermions, two-dimensional (2D) Dirac fermions have been experimentally observed in the AFM phases of $\mathrm{GdSbTe}$ \cite{Hosen18} and $\mathrm{MnBi_2Te_4}$ \cite{Otrokov19,Hao19,Chen19,Li19} as surface states.

\begin{figure*}[ht]
\includegraphics[width=0.9\textwidth]{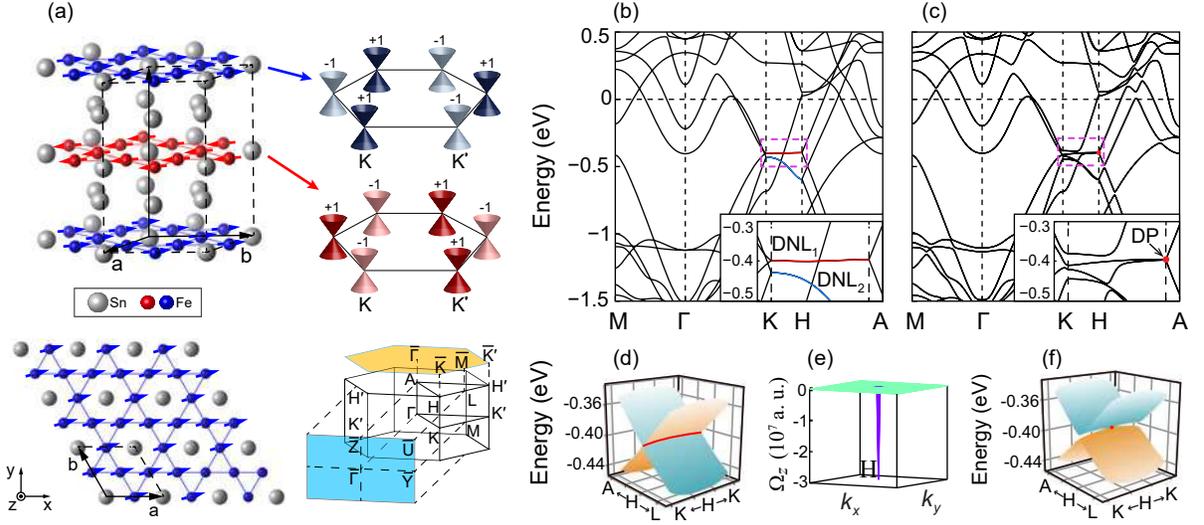}
\caption{\label{figure:1} Electronic band structure of antiferromagnetic FeSn. (a) Rhombohedral unit cell of FeSn, where the small and large circles represent Fe and Sn atoms, respectively. The magnetic moments of Fe atoms are coupled with in-plane FM order and out-of-plane AFM order. Each $\mathrm{Fe_3Sn}$ kagome layer generates 2D Weyl-like states with different chiralities $\chi$ = $\pm$1. Top view of $\mathrm{Fe_3Sn}$ kagome layer and the three-dimensional Brillouin zone (BZ) with its projected surface BZ are also given. The arrows represent the magnetic moments pointing along the [3.732, 1, 0] direction. The $x$, $y$, and $z$ axes point along the [100], [120], and [001] directions, respectively. (b),(c) Calculated band structure without and with SOC, respectively. (d),(e) Two crossing bands of $\mathrm{DNL_1}$ and the Berry curvature component $\mathrm{\Omega_z}$ around the H point in (b). (f) Two crossing bands around the Dirac point (DP) in (c). The Fermi energy $E_{\rm{F}}$ is set at zero energy.}
\end{figure*}

The kagome lattices have attracted much attention for the emergence of both linearly dispersive bands and dispersionless flatbands \cite{Mielke91,Mielke92,Kuroda17,Ye18,Lin18,Liu18}. In $\mathrm{Fe_3Sn_2}$ containing ferromagnetic (FM) kagome lattices, an angle-resolved photoemission spectroscopy (ARPES) experiment reported the existence of massive Weyl states \cite{Ye18}. Meanwhile, in $\mathrm{Mn_3Sn}$ containing AFM kagome lattices \cite{Kuroda17} and $\mathrm{Co_3Sn_2S_2}$ containing FM kagome lattices \cite{Liu18}, massless Weyl states were experimentally measured. Here, the combined density-functional theory (DFT) and ARPES study demonstrates the presence of stable 3D Dirac states in the AFM FeSn system with \emph{PT} symmetry. We further reveal the strong correlations between topological natures and spin degree of freedom via breaking the \emph{PT} or $S_{\rm{2z}}$ symmetry.

Figure 1(a) shows the structure of FeSn consisting of stacked $\mathrm{Fe_3Sn}$ kagome and Sn honeycomb layers, where the Fe atoms form a kagome lattice. Earlier neutron diffraction and M$\mathrm{\ddot{o}}$ssbauer experiments \cite{Yamaguchi67,Kulshreshtha81,Kulshreshtha81} reported that the adjacent $\mathrm{Fe_3Sn}$ layers are antiferromagnetically coupled with each other, and the magnetization in each $\mathrm{Fe_3Sn}$ kagome lattice was measured to be parallel to the [3.732,1,0] direction \cite{Kulshreshtha81}. This AFM kagome system breaks each of the \emph{P} and \emph{T} symmetries but respects the combined \emph{PT} symmetry, providing an ideal platform to realize massless Dirac fermions, as demonstrated below.

We performed first-principles DFT calculations for the band structure of bulk FeSn. The calculated band structure in the absence of SOC shows that all the spin-up and spin-down bands are degenerate with each other because of \emph{PT} symmetry. As shown in Fig. 1(b), there are two Dirac nodal lines (DNLs) $\mathrm{DNL_1}$ and $\mathrm{DNL_2}$ along the K-H line around -0.4 eV. Since the crystalline symmetry of FeSn belongs to the space group P6/mmm (No. 191) with the point group $D_{6h}$, two additional DNLs also exist along the K'-H' line. It is noted that along the K-H and K'-H' lines, the system has the three-fold rotation symmetry $C_{3z}$ around the $z$ axis and \emph{PT} symmetry, thereby preserving the four-fold degeneracy of the DNLs in the absence of SOC \cite{Tang16,Yang15}. In Fig. 1(d) and 1(e), the band-touching point H along $\mathrm{DNL_1}$ exhibits a singularity of the Berry curvature. Further, the topological $Z_2$ index, defined as $\zeta_1$=$\frac{1}{\pi}\oint_{C}\,dk \cdot A(k)$  where $A(k)=-i<u_k|\partial_k|u_k>$ is the Berry connection of Bloch bands, along a closed loop encircling each DNL is calculated to be $\pm$1, indicating that the DNLs are stable against perturbations without breaking the $C_{3z}$ and \emph{PT} symmetries.

\begin{figure}[ht]
\includegraphics[width=\columnwidth]{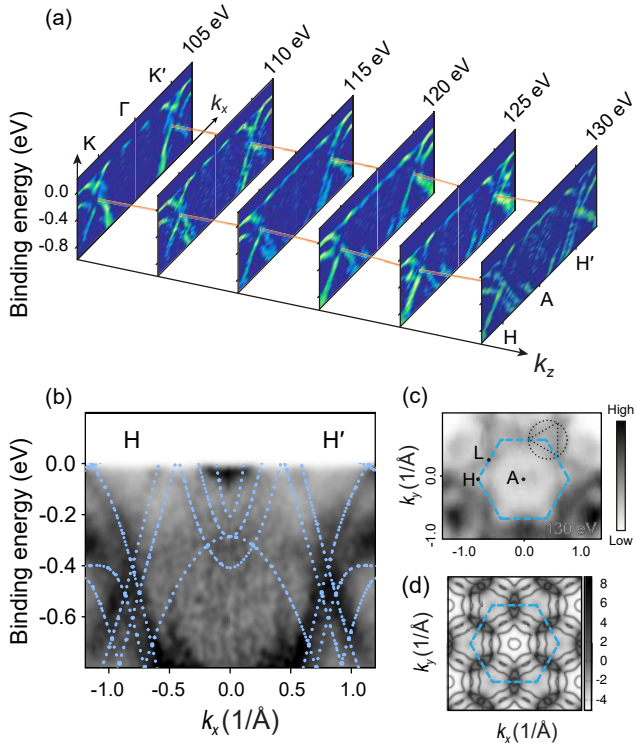}
\caption{\label{figure:2} ARPES spectra of bulk FeSn. (a) Second derivatives of ARPES data between the K-$\Gamma$-K' ($k_z$ = 0) and H-A-H' ($k_z$ = $\pi/c$) lines with changing photon energy from 105 eV to 130 eV, all the data are symmetrized with respect to $k_x$ = 0. (b) Highly resolved ARPES data along H-A-H' line with a photon energy of 130 eV. Here, the ARPES data are not symmetrized with respect to $k_x$ = 0, and the calculated band structure with including SOC is superimposed by the light blue dash lines. (c),(d) Bulk Fermi surface obtained from ARPES measurement and DFT bands of Fig. 1(c).}
\end{figure}

Meanwhile, the bulk band structure with including SOC reveals that the four-fold degeneracy at the band-crossing points along the K-H (K'-H') line is lifted, but is still preserved at the H (H') point [Fig. 1(c) and 1(f)]. Consequently, the SOC gaps of less than $\sim$30 meV appear along the DNLs, and the massless Dirac points exist at the H and H' points. Since magnetization is parallel to the [3.732,1,0] direction as observed by experiment \cite{Kulshreshtha81}, the system has the magnetic space group $\mathrm{P2_1}$/m (No. 11.50), which contains the two-fold screw rotation symmetry ($S_{\rm{2z}}$) along the $z$ axis and \emph{PT} symmetry. Such a non-symmorphic symmetry of $S_{\rm{2z}}$ (equivalent to the combination of $C_{\rm{2z}}$ and a half translation along the $z$ direction) protects the Dirac points at the Brillouin zone boundary points H and H' \cite{Armitage18,Yang14,Tang16}. Thus, the AFM FeSn hosts the massless Dirac states even in the presence of SOC. We further find the existence of the Fermi arcs surface states on the (100) surface, which connect the two Dirac points (Fig. S1 in the Supplemental Material \cite{SM}). Specifically, we find that the spin texture represents a unique spin-momentum locking property of the nontrivial topological surface states. Therefore, the AFM FeSn has not only 3D massless Dirac fermions in bulk but also topologically protected surface states on the (100) surface.

To verify our theoretical prediction of massless Dirac fermions in the AFM FeSn, we have grown high-quality FeSn single crystals to characterize their electronic structures [for the structural characterization results, see Fig. S2(a)-(c) in the Supplemental Material \cite{SM}]. The measured magnetization as a function of temperature shows a peak at  $T_{\rm{N}} \approx$ 366 K [see Fig. S2(d) in the Supplemental Material \cite{SM}], consistent with previous results \cite{Yamaguchi67,Kulshreshtha81}. This AFM structure is further supported by our DFT calculations (see Fig. S3 and more details in the Supplemental Material \cite{SM}) as well as recent experimental measurements of neutron scattering, magnetic susceptibility, and magnetization \cite{Sales19}.

The low-energy electronic structure of the cleaved (001) surface was measured by ARPES. We employed large photon energies ranging from 90 to 160 eV to acquire the ARPES data between the K-$\Gamma$-K' and H-A-H' lines, which enable us to examine the bulk states. Figure 2(a) shows the ARPES data of photon energies between 105 and 130 eV, representing the energy bands from $k_{\rm{z}}$ = 0 to $k_{\rm{z}}$ = $\pi/c$ (see Fig. S4, Fig. S5 and more details in the Supplemental Material \cite{SM}). We find that the apparent crossings of two linearly dispersive bands appear at the K and K' points around 0.4 eV below the Fermi energy $E_{\rm{F}}$ (the red arrows in Fig. S5 in the Supplemental Material \cite{SM}). Remarkably, such crossing positions are nearly identical with increasing photon energy, indicating the presence of the DNLs along the K-H and K'-H' lines, as predicted by DFT calculations [Fig. 1(c)]. Figure 2(b) shows the ARPES spectrum along H-A-H' line, from which no obvious gap opening is detected at the Dirac points around the H and H' points. However, we cannot unambiguously identify whether these Dirac points are gapless or not due to the limited spectroscopic resolution ($\sim$10 meV).

\begin{figure}
\includegraphics[width=\columnwidth]{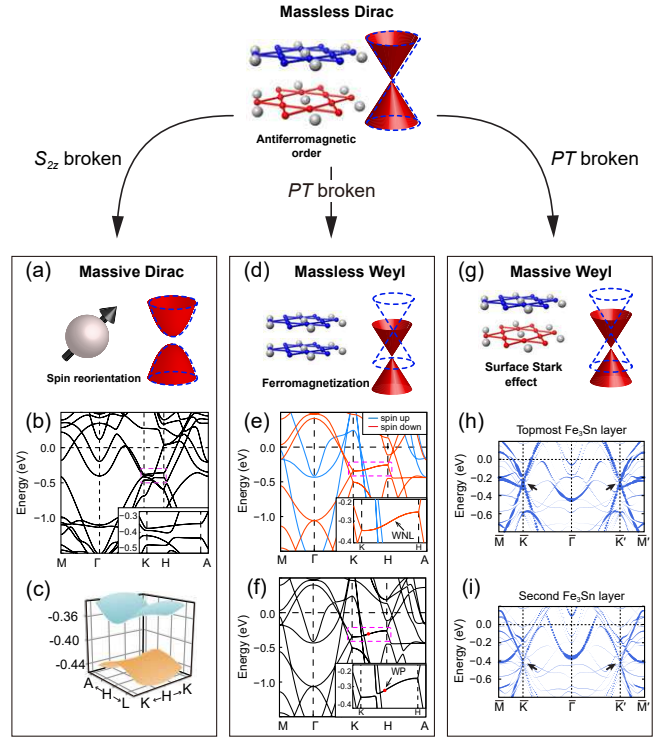}
\caption{\label{figure:3} Topological phase transitions in FeSn. (a) Transition from massless Dirac to massive Dirac fermions via breaking $S_{\rm{2z}}$ symmetry by spin reorientation. Note that the manipulation of the N$\mathrm{\acute{e}}$el spin orientation can tune the magnetic symmetry to break $S_{\rm{2z}}$ symmetry and generate non-zero mass in the originally massless Dirac bands. (b) Calculated band structures with SOC, where the spin orientations are along the [001] directions. (c) Two bands around the H point in (b). (d) Transition from massless Dirac to massless Weyl fermions via breaking \emph{PT} symmetry by ferromagnetization. (e),(f) The calculated band structures of the FM phase of bulk FeSn, which show the existences of massless Weyl nodal line (without SOC) and Weyl point (with SOC). (g) Transition from massless Dirac to massive Weyl fermions via breaking \emph{PT} symmetry by surface stark effect. (h),(i) Projected band structures of the FeSn(001) surface on the topmost (h) and second $\mathrm{Fe_3Sn}$ layers (i), obtained with SOC. Here, the radii of solid circles are proportional to the weights of the corresponding local density of states.}
\end{figure}

Furthermore, when the theoretically calculated bands overlap with the ARPES intensity along the H-A-H' line, both data are in good agreement in a large energy window [Fig. 2(b)]. The relatively weaker ARPES intensity [Fig. 2(b)] in the right (left) side of the upper part of the Dirac bands at the H (H') point is likely due to the spin-selective effect \cite{Xie14,Jozwiak13,Gierz11}, as observed in a recent ARPES experiment \cite{Kang20}. In particular, the bulk Fermi surface measured with a 130-eV photon energy [Fig. 2(c)] exhibits a triangular-like pocket centered at the H point as well as a circular pocket outside the triangular-like electron pocket, in good agreement with that [Fig. 2(d)] obtained from the SOC-included DFT calculation.

The symmetry protection of Dirac points in bulk FeSn can be manipulated via spin reorientation that can break $S_{\rm{2z}}$ symmetry [Fig. 3(a)]. Our SOC-included DFT calculations show that, when the spin orientations are still in the (001) plane, i.e., along the [100] and [210] directions, the massless Dirac points at the H and H' points are preserved (see Fig. S7 in the Supplemental Material \cite{SM}). However, as shown in Fig. 3(b) and 3(c), the calculated band structure with the [001] spin orientation perpendicular to the (001) plane opens a gap of $\sim$70 meV at the H and H' points, due to breaking of the non-symmorphic symmetry $S_{\rm{2z}}$. Because of a tiny magnetic anisotropy energy of $\sim$0.03 meV/unit-cell in FeSn crystal, we anticipate that external perturbations, e.g., spin-orbit torque, can readily manipulate the spin orientations to control the mass of Dirac fermions \cite{Smejkal17,Smejkal18}.

Further, the topological quantum states of FeSn can also be tuned via breaking the \emph{PT} symmetry. For instance, turning the interlayer coupling from AFM to FM can induce the splitting of the Dirac point into two two-fold degenerate Weyl points [Fig. 3(d)]. Our DFT calculations confirm that the DNL along K-H line is transformed into a Weyl nodal line (WNL) in the absence of SOC [Fig. 3(e)]. When SOC is included, the WNL open gaps, but a gapless Weyl point is still present [Fig. 3(f)].

It is thus demonstrated that the spin degree of freedom in the present AFM kagome system serves as a new knob to effectively tune the topological behaviors. As schematically shown in Fig. 1(a), each Dirac point can be considered to be composed of a pair of Weyl points arising from antiferromagnetically coupled adjacent FM $\mathrm{Fe_3Sn}$ kagome layers. However, on the FeSn(001) surface breaking the \emph{PT} symmetry, the paired Weyl points are no longer degenerate in energy due to the surface potential difference between the topmost and second $\mathrm{Fe_3Sn}$ kagome layers [Fig. 3(g)]. The surface potential is recovered to the bulk one even at the second $\mathrm{Fe_3Sn}$ layer, indicating that the surface-induced Stark effect occurs only at the topmost $\mathrm{Fe_3Sn}$ layer (see Fig. S8 in the Supplemental Material \cite{SM}). Our calculated band structures of FeSn(001) show that the surface-induced Stark effect splits each Dirac band into two spin-polarized nondegenerate Weyl bands, which reside in neighboring $\mathrm{Fe_3Sn}$ surface and subsurface layers: i.e., one species located at the topmost $\mathrm{Fe_3Sn}$ layer shifts toward a higher energy up to $\sim$-0.2 eV at the $\mathrm{\overline{K}}$ and $\mathrm{\overline{K}}$' points [Fig. 3(h)], while the other species located at the second $\mathrm{Fe_3Sn}$ layer still remains around -0.4 eV [Fig. 3(i)]. Such paired but real-space separated Weyl-like bands originating from the topmost (second) $\mathrm{Fe_3Sn}$ kagome layer have a small gap opening of $\sim$5 (1) meV with including SOC \cite{You19,Wu19}, representing 2D massive Weyl-like bands. We note that the magnetic moment in the topmost $\mathrm{Fe_3Sn}$ layer is 2.16 $\mathrm{\mu_B}$ per Fe atom, slightly larger than that (1.96 $\mathrm{\mu_B}$) in bulk $\mathrm{Fe_3Sn}$ layers (see Fig. S9 in the Supplemental Material \cite{SM}). Interestingly, the bulk Dirac and surface Weyl states are composed of the $d_{xy}$ and $d_{x^2-y^2}$ orbitals (see Fig. S10 in the Supplemental Material \cite{SM}), while the FM instability in each $\mathrm{Fe_3Sn}$ layer originates from the flatbands consisting of the $d_{xz}$ and $d_{yz}$ orbitals (see Fig. S3 in the Supplemental Material \cite{SM}).

\begin{figure}[htb]
\includegraphics[width=0.9\columnwidth]{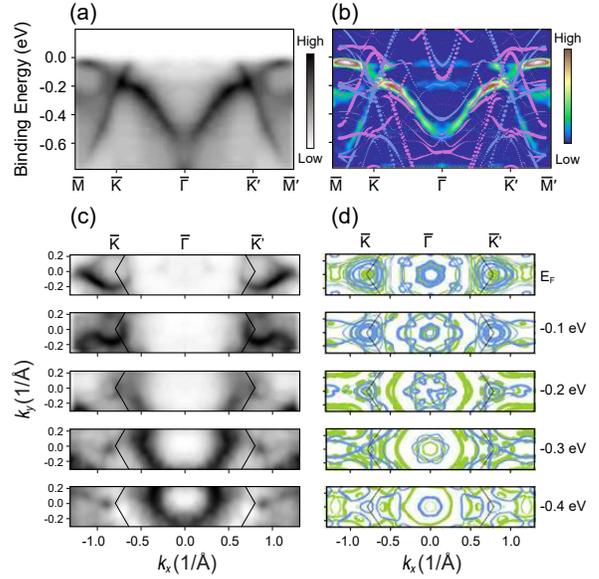}
\caption{\label{figure:4} Electronic band structure of the FeSn (001) surface. (a),(b) ARPES data (a) with a 35 eV-photon energy and their second derivatives (b) along the $\mathrm{\overline{K}}$-$\mathrm{\overline{\Gamma}}$-$\mathrm{\overline{K}}$' line. In (b), the projected DFT band structures of the Sn-terminated (blue color circles) and $\mathrm{Fe_3Sn}$-terminated (purple color circles) surfaces are overlapped. Here, the radii of circles are proportional to the local density of states projected onto the topmost Sn and $\mathrm{Fe_3Sn}$ surface layers. (c) Constant energy surface with respect to chemical potential, obtained by ARPES with a 35 eV-photon energy. The ARPES data are symmetrized with respect to $k_x$ = 0. (d) Projected DFT constant energy surfaces of the Sn-terminated (blue color circles) and $\mathrm{Fe_3Sn}$-terminated (green color circles) surfaces.}
\end{figure}

To verify the emergence of 2D Weyl-like states at surface, we acquired the ARPES data with a low photon energy of 35 eV (see Fig. S11 and more discussions in the Supplemental Material \cite{SM}). Strikingly, as shown in Fig. 4(a) and 4(b), the ARPES data along the $\mathrm{\overline{K}}$-$\mathrm{\overline{\Gamma}}$-$\mathrm{\overline{K}}$' line reveal the crossings of two linearly dispersive bands at the $\mathrm{\overline{K}}$ and $\mathrm{\overline{K}}$' points only around -0.2 eV, differing from the bulk Dirac points around -0.4 eV [Fig. S5 in the Supplemental Material \cite{SM}]. Although our scanning tunneling microscope measurements identified the surface structure in terms of the Sn-terminated surface (see Fig. S12 and more details in the Supplemental Material \cite{SM}), the ARPES probes a large surface area which is likely composed of two different structures with the Sn-terminated and $\mathrm{Fe_3Sn}$-teminated surfaces. As shown in Fig. 4(b), the main features of the APRES data can be reproduced from the DFT results obtained from these two differently terminated surfaces. For instance, the observed linearly dispersive bands fit well with two crossing DFT bands at the $\mathrm{\overline{K}}$ or $\mathrm{\overline{K}}$' point around -0.2 eV obtained from the Sn-terminated surface. Here the absence of ARPES intensity in one side of the lower part of these 2D Weyl-like bands is likely due to the above-mentioned spin-selective effect \cite{Xie14,Jozwiak13,Gierz11}. Meanwhile the parabolic-like ARPES band around the $\mathrm{\overline{\Gamma}}$ point also fits well with the combined multiple DFT bands obtained from the $\mathrm{Fe_3Sn}$-terminated surface. Moreover, the top view of the experimentally observed Fermi surface around the $\mathrm{\overline{K}}$ and $\mathrm{\overline{K}}$' points agree well with the mixed DFT Fermi surfaces of both the Sn-terminated and $\mathrm{Fe_3Sn}$-terminated surfaces [Fig. 4(c) and 4(d)].

In summary, we have demonstrated the presence of 3D Dirac fermions in the AFM FeSn kagome lattices, and further revealed that the versatile topological properties can be entangled with the spin configurations and magnetic symmetries. This realization of Dirac fermions in bulk FeSn further demonstrates the fascinating physics of kagome lattice, together with previous findings of flat bands and Weyl bands \cite{Kuroda17,Ye18,Lin18,Liu18}. By means of further controlling band filling or manipulating the spin degree of freedom, it is highly promising to realize intriguing transport performances that can be utilized for future spintronic devices.

\begin{acknowledgments}
This work was supported by in part by the National Natural Science Foundation of China (Grants No. 11974324, U1832151, and 11804326), the National Key Research and Development Program of China (Grants No. 2017YFA0403600 and 2017YFA0402901), Anhui Initiative in Quantum Information Technologies (Grant No. AHY170000), Hefei Science Center CAS (Grant No. 2018HSC-UE014), the National Research Foundation of Korea (NRF) Grant funded by the Korean Government (Grants No. 2019R1A2C1002975, 2016K1A4A3914691, and 2015M3D1A1070609). The calculations were performed by the KISTI Supercomputing Center through the Strategic Support Program (Program No. KSC-2018-CRE-0063) for the supercomputing application research.\\
\indent $^{\dagger}$Z.L., C.W., P.W. and S.Y. contributed equally to this work.\\

\indent $^{*}$Corresponding authors: cgzeng@ustc.edu.cn, chojh@hanyang.ac.kr, zsun@ustc.edu.cn, lilin@ustc.edu.cn
\end{acknowledgments}

\end{document}